\documentclass[sigconf]{acmart}
\acmConference[ICPC 2024]{46th International Conference on Program Comprehension}{April 2024}{Lisbon, Portugal}
\usepackage{balance} 
\usepackage[skins]{tcolorbox}
\usepackage{multirow}
\usepackage{url}
\usepackage{listings}

\AtBeginDocument{%
  \providecommand\BibTeX{{%
    \normalfont B\kern-0.5em{\scshape i\kern-0.25em b}\kern-0.8em\TeX}}}

\tcbset{
  my box/.style={
    enhanced,
    colframe=#1!80,
    colback=#1!10,
    attach boxed title to top left={xshift=0.2cm, yshift=-0.2cm},
    boxed title style={
      colback=#1!80,
      outer arc=0pt,
      arc=0pt,
      top=0pt,
      bottom=0pt,
    },
  },
}
\newtcolorbox{result-rq}[1]{
  my box=black,
  title=#1,
  boxrule=1.2pt,top=6pt,bottom=3.5pt,left=6pt,right=6pt
}

\copyrightyear{2024}
\acmYear{2024}
\setcopyright{acmlicensed}\acmConference[ICPC '24]{32nd IEEE/ACM International Conference on Program Comprehension}{April 15--16, 2024}{Lisbon, Portugal}
\acmBooktitle{32nd IEEE/ACM International Conference on Program Comprehension (ICPC '24), April 15--16, 2024, Lisbon, Portugal}
\acmDOI{10.1145/3643916.3644401}
\acmISBN{979-8-4007-0586-1/24/04}

\begin{document}

\title[MESIA: Understanding and Leveraging Supplementary Nature of Method-level Comments for Automatic Comment Generation]{MESIA: Understanding and Leveraging Supplementary Nature of Method-level Comments for Automatic Comment Generation}


\author{Xinglu Pan}
\affiliation{%
\institution{SCS Peking University; Key Lab of HCST (PKU), MOE}
  \city{Beijing}
  \country{China}}
\email{bjdxpxl@pku.edu.cn}

\author{Chenxiao Liu}
\affiliation{%
  \institution{SCS Peking University; Key Lab of HCST (PKU), MOE}
  \city{Beijing}
  \country{China}}
\email{jslcx@pku.edu.cn}

\author{Yanzhen Zou}
\affiliation{%
  \institution{SCS Peking University; Key Lab of HCST (PKU), MOE}
  \city{Beijing}
  \country{China}}
\email{zouyz@pku.edu.cn}

\author{Tao Xie}
\affiliation{%
  \institution{SCS Peking University; Key Lab of HCST (PKU), MOE}
  \city{Beijing}
  \country{China}}
\email{taoxie@pku.edu.cn}

\author{Bing Xie}
\authornote{Bing Xie is the corresponding author.}
\affiliation{%
  \institution{SCS Peking University; Key Lab of HCST (PKU), MOE}
  \city{Beijing}
  \country{China}}
\email{xiebing@pku.edu.cn}

\renewcommand{\shortauthors}{Pan et al.}
\begin{abstract}
Code comments are important for developers in program comprehension. In scenarios of comprehending and reusing a method, developers expect code comments to provide supplementary information beyond the method signature. However, the extent of such supplementary information varies a lot in different code comments. In this paper, we raise the awareness of the supplementary nature of method-level comments and propose a new metric named MESIA (\textbf{Me}an \textbf{S}upplementary \textbf{I}nformation \textbf{A}mount) to assess the extent of supplementary information that a code comment can provide. With the MESIA metric, we conduct experiments on a popular code-comment dataset and three common types of neural approaches to generate method-level comments. Our experimental results demonstrate the value of our proposed work with a number of findings. 
(1) Small-MESIA comments occupy around 20\% of the dataset and mostly fall into only the WHAT comment category. 
(2) Being able to provide various kinds of essential information, large-MESIA comments in the dataset are difficult for existing neural approaches to generate. 
(3) We can improve the capability of existing neural approaches to generate large-MESIA comments by reducing the proportion of small-MESIA comments in the training set. 
(4) The retrained model can generate large-MESIA comments that convey essential meaningful supplementary information for methods in the small-MESIA test set, but will get a lower BLEU score in evaluation. 
These findings indicate that with good training data, auto-generated comments can sometimes even surpass human-written reference comments, and having no appropriate ground truth for evaluation is an issue that needs to be addressed by future work on automatic comment generation.

\end{abstract}





\maketitle

\section{Introduction}
As modern software continues to grow in complexity, code comments become increasingly important in program  comprehension~\cite{7997917,woodfield1981effect, kajko2005survey,de2005study,10.1145/3312662.3312710,10.1145/3338906.3342494,6171,10.1145/3377811.3380427}. Code comments are specially essential for methods. In scenarios of comprehending and reusing a method, developers often access only the method signature (without access to the method body or looking into it when accessible), which includes the method's name, parameters, and return type. Although the method signature conveys valuable information for developers, it may be too short~\cite{7884623} to express all essential information needed to comprehend the method. As a result, developers have to turn to code comments to get  supplementary (i.e., additional) information to better comprehend the method.

\begin{figure*}[t]
  \centering  \includegraphics[height=4.5cm,width=12cm]{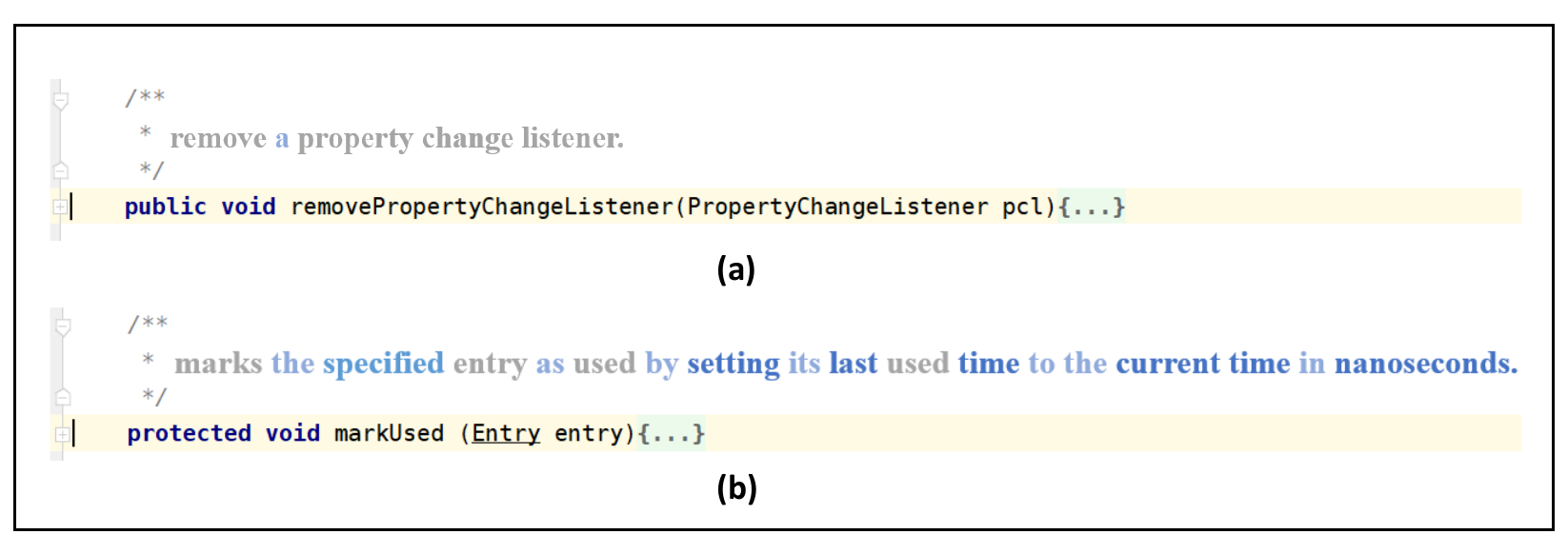} 
  \caption{Different code comments that provide different extents of supplementary information.}
  \label{fig:differentComments}
  \Description{}
\end{figure*}

Developers typically follow a \textbf{hybrid mode} for reading the method-level comment to comprehend a method where developers read both the method signature and the comment. As mentioned earlier, because developers have digested the method signature, they expect the comment to provide more supplementary information beyond the method signature~\cite{10.1145/3510003.3510152,PhilosophyOfSoftware}. According to a recent study~\cite{10.1145/3510003.3510152} on professional developers' expectations on code comments, developers state that ``code is the best documentation'' and code comments shall provide supplementary information beyond what can be easily obtained from code~\cite{PhilosophyOfSoftware}, especially the method signature.


Although developers expect code comments to provide supplementary information, we have observed that the extent of supplementary information varies a lot in different code comments as shown in the following two examples\footnote{Both examples are from an existing code-comment dataset named  TL-CodeSum~\cite{TL-CodeSum} studied in this paper.}. (1)  Figure~\ref{fig:differentComments}(a) shows a code comment that provides little supplementary information. The comment of the method \emph{``removePropertyChangeListener''} is \emph{``remove a property change listener''}, which provides only an \emph{``a''} besides the content that can be directly obtained from the method name. (2) Compared with the preceding example, the comment of the method \emph{``markUsed''} in Figure~\ref{fig:differentComments}(b) can provide a larger extent of supplementary information. Here a large part of the comment content (such as \emph{``by setting its last used time to the current time in nanoseconds''}) is supplementary information that cannot be directly obtained from the method signature.

Toward supporting the hybrid mode of reading method-level comments to comprehend a method, in this paper, we propose a new metric named MESIA (\textbf{Me}an \textbf{S}upplementary \textbf{I}nformation \textbf{A}mount) to quantify the extent of supplementary information that a method-level comment can provide beyond the information reflected by the method signature. The insight of the MESIA metric comes from the self-information given by the Shannon information theory~\cite{6773024}. It quantifies the information amount of a word by the surprise level of its particular outcome. In particular, to compute MESIA, we first calculate the information amount of each comment word by the possibility of the word appearing in all comments in a dataset. We then  calculate the supplementary extent of a method comment by considering the information amount of all words in the comment, all words in its related method signature, and the length of the comment, respectively.

We apply the MESIA metric on a popular code-comment dataset named TL-CodeSum~\cite{TL-CodeSum}. Our empirical results show that MESIA is consistent with manual assessment of the relative supplementary extent of the code comment beyond the method signature of a method. We investigate the distribution of various kinds of code comments in different value ranges of MESIA to understand the supplementary nature of method-level comments. The results show that small-MESIA-value (in short as small-MESIA) comments  (with less than 3 MESIA value), occupying around 20\% of the dataset, mostly fall into only the WHAT comment category (describing the functionality)~\cite{10.1145/3434280} while the large-MESIA-value (in short as large-MESIA) comments (with greater than 6 MESIA value) can widely cover all six comment categories.

We further apply our proposed MESIA metric to three common types of neural approaches~\cite{iyer2016summarizing,ahmad-etal-2020-transformer,wang-etal-2021-codet5} to generate code comments with different MESIA values. By training a neural network model with pairs of a code snippet (usually in the method-level granularity) and its comment, the model can learn to translate a code snippet to its comment. 
Our results show that the existing neural approaches are far from satisfactory when used to generate large-MESIA comments, reflecting challenging and important cases for future research efforts to tackle. Subsequently, we explore strategies that leverage MESIA to improve the generation of large-MESIA comments. The results indicate that we can improve the capability of existing neural approaches to generate large-MESIA comments by reducing the proportion of small-MESIA comments in the training set.  In addition, the retrained model can generate large-MESIA comments that convey essential meaningful supplementary information for methods in the small-MESIA test set, but will get a lower BLEU score in evaluation. These findings indicate that with good training data, auto-generated comments can sometimes even surpass human-written reference comments, and having no appropriate ground truth for evaluation is an issue that needs to be addressed by future work on automatic comment generation. Finally, we discuss the challenges of generating large-MESIA comments for future work on automatic comment generation to tackle.

Our work raises the awareness of developers' expectations for the supplementary information of method-level comments in a hybrid mode of reading the method-level comment to comprehend a method. Our experimental results reveal the influence of the supplementary nature of method-level comments on automatic comment generation. The results can guide the usage of data in future work to effectively generate comments that support developers' hybrid mode of comprehending a method. 

In summary, this paper makes the following main contributions:
\begin{itemize}
\item A new metric named MESIA to assess the extent of supplementary information that a code comment for a method can provide beyond the method signature.

\item
An empirical investigation of the MESIA metric on a popular code-comment dataset, showing that small-MESIA comments occupy around 20\% of the dataset and mostly fall into only the WHAT comment category, while large-MESIA comments can provide various kinds of essential information about methods.

\item
An evaluation upon three common types of neural approaches for generating method-level comments, showing that the existing approaches are far from satisfactory in generating the large-MESIA comments.

\item
A strategy to improve automatic generation of large-MESIA comments with existing neural approaches by reducing the proportion of small-MESIA comments in the training set. 
\end{itemize}

The rest of the paper is organized as follows. Section~\ref{sec:motivation} describes the motivation and challenges addressed by our work. Section~\ref{sec:Definition} illustrates our proposed MESIA metric. Section~\ref{sec:study design} describes the study design of our research. Section~\ref{sec:result} presents the study results. Section~\ref{sec:discussion} discusses the implications and threats to validity. Section~\ref{sec:relatedwork} describes the related work. Finally, Section~\ref{sec:conclusion} concludes this paper.

\section{Motivation And Challenges}
\label{sec:motivation}

In this section, we explain our motivation and challenges to be addressed by our work.

Our motivation comes from the observation that in the hybrid mode of reading the method-level comment for a method, developers expect the  comment to provide more supplementary information beyond the method signature. So besides the existing widely considered aspects in assessing code comments~\cite{10.1016/j.jss.2022.111515} (such as accuracy~\cite{sridhara2011generating,mcburney2014automatic}, adequacy~\cite{10.1109/ICSE48619.2023.00073}, and naturalness~\cite{iyer2016summarizing}), we also need a metric to assess such supplementary extent of the code comment beyond the method signature. The assessment should concern, but not limit to, the following three factors.

First, we need to consider how to evaluate the information amount of a given code comment. The information amount of a code comment relies on the information amount of its words, which varies a lot in the same dataset. Take the comment \emph{``marks the specified entry as used by setting its last used time to the current time in nanoseconds''} as an example, words such as ``the'',  ``specified'', and ``to'' are often used in code comments of the dataset and they can provide little information. Words such as ``setting'', ``current'', and  ``time'' are less often used and they provide more information to developers.

Second, we need to consider the supplementary information amount of the given code comment beyond the method signature. Code is the best documentation~\cite{10.1145/3510003.3510152} and developers can directly get some information from the method signature~\cite{10.1145/3180155.3180176}, especially when it follows a good naming convention. For example, given the method signature \emph{``markUsed(Entry entry)''}, developers can directly know that this method will mark the entry used. It is necessary to reduce the impact of the method signature in the metric.

Third, we need to consider the length (how many words) of the given code comment. Because code comments in different methods inherently have different lengths, the absolute supplementary information amount of the code comments can vary greatly. What we care about is, under the comment length, whether the comment has provided a relatively large extent of supplementary information. Therefore, we should consider the length of the comment to measure its supplementary extent so that the result is comparable across different comment entries.

\begin{figure}[t]
  \centering
  \includegraphics[height=5cm,width=8.5cm]{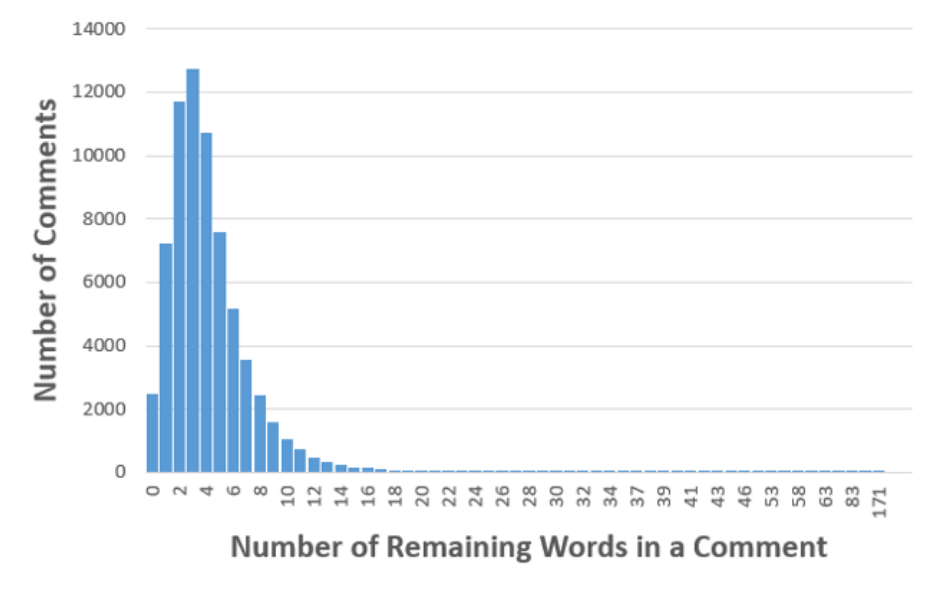}
  \caption{An analysis of the number of words that remain in a method comment after removing stop words and words in the method's split signature.}
  \label{fig:analysis1}
  \Description{}
\end{figure}

\begin{figure}[t]
  \centering
  \includegraphics[height=5cm,width=8.5cm]{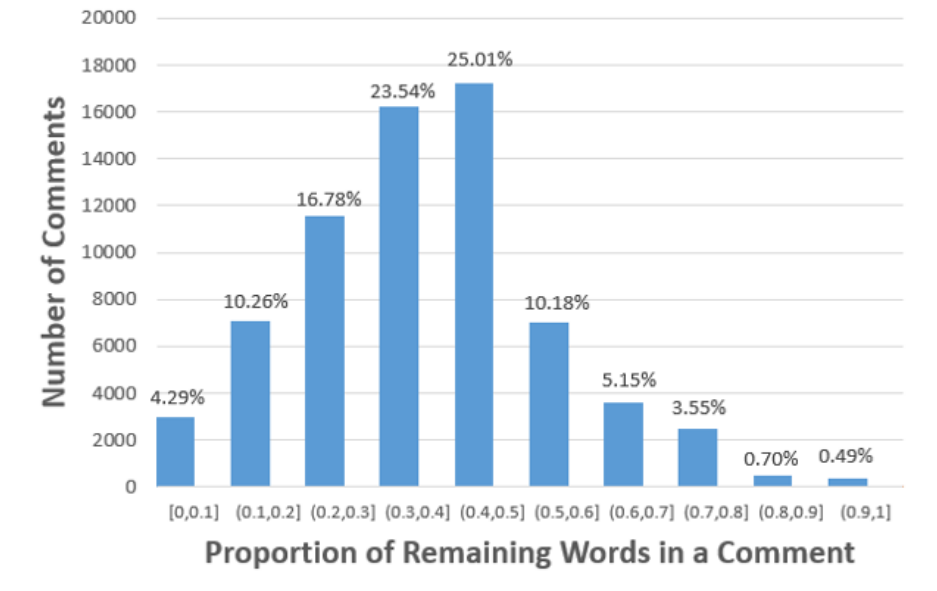}
  \caption{An analysis of the proportion of words that remain in a method comment after removing stop words and words in the method's split signature.}
  \label{fig:analysis2}
  \Description{}
\end{figure}

To get an initial understanding of these three factors, we conduct a two-step analysis on an existing dataset~\cite{TL-CodeSum}. This dataset is widely used by existing neural approaches for generating method-level comments. First, we calculate the number of words that remain in the given code comment after we remove the stop words and the words appearing in its method signature. Second, we calculate the proportion of the remaining words in the given code comment by further considering its original length. In the analysis, we split the words in the method signature. The split comprehensively considers the camel case spelling and snake case spelling. Then we stem these split words and words in the given code comment to a normalized form for comparison. For example, the comment of the method \emph{``markUsed''} is \emph{``marks the specified entry as used by setting its last used time to the current time in nanoseconds.''}. After removing the stop words (``the'', ``as'', ``by'', ``its'', ``to'', ``in''), the words in the split method name (``marks'', ``used''), and the words in the split parameter's name (``entry''), seven words (``specified'', ``setting'', ``last'', ``time'', ``current'', ``time'', ``nanoseconds'') remain. The proportion of the remaining words in the comment is 7/18 $\approx$ 38.9\%.

From Figure~\ref{fig:analysis1}, we can see that after we remove the stop words and the words in the method's split signature, the number of remaining words in a code comment mainly focuses on [0,8] and there is a long tail. So referring to the results of Figure~\ref{fig:analysis2}, we need to evaluate the extent of supplementary information for different code comments based on the proportion of their remaining words. However, the specific calculation is still a great challenge because different remaining words can provide different amount of information.

\section{Preliminary and Definition}
\label{sec:Definition}
As described in the preceding section, we want to evaluate the extent of supplementary information in a method comment beyond its method signature. We borrow the idea from Shannon information theory and  propose a new metric named MESIA (\textbf{Me}an \textbf{S}upplementary \textbf{I}nformation \textbf{A}mount) to assess code comments.

In 1948, Shannon~\cite{6773024} derived a metric of information content named the self-information of a message x:
\begin{equation}
    I(x) = -log(p(x))
\end{equation}
Here $p(x)$ is the probability that message x is chosen from all possible choices in the message space X. The self-information can be interpreted as quantifying the surprise level of a particular outcome. When a message with less appearing possibility occurs, it brings more information.

To calculate the information amount of a given code comment, we first follow the self-information to measure the information amount for each word $w$ in the given code comment as 
\begin{equation}
    I(w) = -log(p(w))
\end{equation}
Here $p(w)$ represents the possibility of $w$ appearing in code comments. The basic idea behind this equation is that if a word is more likely to appear in code comments, the less extra information it can bring to developers. 

Thus we calculate the information amount for the given code comment by summing up the information amount of each word in the comment. We define a code comment in a dataset as $C = <w_{1},w_{2},...,w_{n}>$. All code comments in the  dataset are defined as $ Comments = \{C_{1},C_{2},...,C_{m}\}$.
All words used in comments of the dataset are defined as $W = \bigcup\limits_{i=1}^m \{w|w \in C_{i}\}$. We use the frequency of each word $x$ (referred to as $freq(x)$) in $W$ to calculate the possibility for each word $w$ in the given code comment. 
\begin{equation}
p(w|W) = freq(w)/\sum_{x\in W}^{}(freq(x))
\end{equation}
We then define the information amount of a comment C in the dataset as 
\begin{equation}
    I(C|W) = -\sum_{w \in C}log(p(w|W))
\end{equation}


Second, we further consider the method signature in the metric. Code comments serve as a way to provide supplementary information in program comprehension. If developers can easily obtain the comment's content from the method signature, then the comment brings little supplementary information. Therefore, we define the supplementary information amount of a comment C beyond its method signature as
\begin{equation}
I(C|Code,W) = -\sum_{w \in C}log(p(w|Code,W))
\end{equation}
Here the ``$Code$'' in the formula refers to the words in the method's split signature. We consider the method signature in the metric as presented in the following formulas
\begin{equation}
p(w|Code,W) =
\begin{cases}
1& w \in Code\\
p(w|W)  & w \notin Code
\end{cases}
\end{equation}
When $w$ appears in the method's split signature, we set $p(w|Code,W)$ to 1. If $w$ does not appear in the method's split signature, we set $p(w|Code,W)$ to $p(w|W)$.  We split the method signature into words according to two heuristic rules: (1) camel case spelling and (2) snake case spelling. We stem these split words and words in the given code comment to a normalized form for comparison. The details have been introduced in Section~\ref{sec:motivation}.

Finally, the length of the given comment must be considered because we want to evaluate the extent of supplementary information for code comments. Therefore, we define the MESIA (\textbf{Me}an \textbf{S}upplementary \textbf{I}nformation \textbf{A}mount) of a comment C as 
\begin{equation}
\begin{split}
MESIA(C) &= I(C|Code,W)/len(C) \\
        &= -\sum_{w \in C}log(p(w|Code,W))/len(C)
\end{split}
\end{equation}

In the following sections, we use the MESIA metric to assess code comments in an existing dataset and conduct several experiments.

\section{Study Design}
\label{sec:study design}

In this section, we present the study design for investigating the MESIA metric toward understanding the supplementary nature of method-level comments in an existing dataset and understanding the supplementary nature's influence on three common types of neural approaches to generate method-level comments.
We mainly aim to answer three research questions. 

\subsection{Research Questions}
\textbf{RQ1. How well does MESIA reflect the relative extent of supplementary information in method-level comments?}

Answering this research question helps investigate whether MESIA is consistent with manual assessment on the relative supplementary extent of code comments. We also want to find out the distribution of various kinds of code comments in different value ranges of MESIA to better understand the supplementary nature of method-level comments.

\textbf{RQ2. What is the capability of existing neural approaches to generate code comments with different MESIA values?}

In this paper, we raise the awareness of the supplementary nature of method-level comments. Answering this research question helps investigate the capability of existing neural approaches to generate code comments with various extents of supplementary information. 

\textbf{RQ3. How well can MESIA be used to improve existing neural approaches to generate large-MESIA comments?}

The neural approaches for generating code comments require a large amount of code-comment data. The training data can substantially influence the comment-generation result. Answering this research question helps investigate whether we can help existing neural approaches to generate code comments that better support developers in a hybrid mode of reading method-level comments after we leverage MESIA to refine the training data.

\subsection{Study Setup}
The background of our study is the recent increasing popularity of neural approaches~\cite{iyer2016summarizing,ahmad-etal-2020-transformer,hu2018deep,10.1145/3368089.3417926,wan2018improving,9284039,10.1145/3324884.3416578,9678882,9678724,alon2018code2seq,leclair2019neural} to generate method-level comments. These approaches typically consist of three steps. First, these neural approaches construct a dataset consisting of pairs of a method and its comment. The dataset is divided into three parts: training, validation, and test sets. Second, using the training and validation sets, the neural approaches train a neural network model, such as the sequence-to-sequence (seq2seq) model~\cite{10.5555/2969033.2969173} or the Transformer model~\cite{10.5555/3295222.3295349}, to learn the alignment of a method and its comment. Third, using the neural network model, the neural approaches generate a comment for the given method in test set and evaluate the generated comment by calculating its similarity with the corresponding reference comment via metrics such as BLEU~\cite{papineni2002bleu}.

\subsubsection{Dataset}

We use a popular code-comment dataset named TL-CodeSum~\cite{TL-CodeSum} provided by Hu et al.~\cite{ijcai2018-314} to conduct our study. This dataset originally contains about 87K pairs of a method and its comment collected from more than 9K open-source Java projects created from 2015 to 2016 with at least 20 stars in GitHub. Hu et al.  extract the methods and their corresponding Javadoc comments, and take the first sentence of the comments and use some heuristic rules to remove some methods such as  setter, getter, and test methods.

Recently, much work~\cite{10.1145/3510003.3510060,10.1145/3540250.3549145} has pointed out the quality issues of existing  datasets of code comments. Specifically, Shi et al.~\cite{10.1145/3540250.3549145} point out that noise intensively exists in existing datasets and the TL-CodeSum dataset contains a certain percentage of data duplication. The duplication of data will influence the evaluation of existing approaches because the generalization ability of the model cannot be well assessed. Therefore, we use the denoised version~\cite{Denoised-TL-codesum} of the TL-CodeSum dataset provided by Shi et al.~\cite{10.1145/3540250.3549145} and further deduplicate it to conduct our study. The deduplication process is as follows: (1) We compare code-comment samples in the training set and remove duplicated samples so that each sample in the training set is unique. (2) Following (1), we deduplicate the validation set and further remove samples of the validation set that appear in the training set. (3) Following (1), we deduplicate the test set and further remove samples in the test set that appear in the validation set or the training set. Table~\ref{tab:javadata} shows the detail of the dataset used in our study.

\begin{table}[t]
  \caption{The TL-CodeSum Dataset.}
    \label{tab:javadata}
  \centering
  \begin{tabular}{c|ccc}
    \hline
      & \#Training Pairs & \#Validation Pairs & \#Test Pairs \\
    \hline
     Original & 69708 & 8714 & 8714 \\
     Denoised & 53597 & 7562 & 7584 \\
     Deduplicated &53506 & 6040 & 5905\\
     \hline
  \end{tabular}
\end{table}

\subsubsection{Neural Approaches}

We use MESIA to investigate three common types of neural approaches~\cite{iyer2016summarizing,ahmad-etal-2020-transformer,wang-etal-2021-codet5} for generating method-level comments and study their results. The first approach uses the  seq2seq model~\cite{10.5555/2969033.2969173}, which contains two  layers of the LSTM encoder and one layer of the LSTM decoder. The second approach uses the Transformer model~\cite{10.5555/3295222.3295349}, which uses the multi-head attention layer and contains six layers of the encoder and six layers of the decoder. The third approach is a pre-trained code model named CodeT5~\cite{wang-etal-2021-codet5}, which can be fine-tuned in pairs of code-comment data to generate code comments. These are three representative approaches and other sophisticated approaches~\cite{hu2018deep,10.1145/3368089.3417926,wan2018improving,9284039,10.1145/3324884.3416578,9678882,9678724,alon2018code2seq,leclair2019neural} are more or less based on ideas from these three approaches. Thus, by studying these three approaches, we can get a basic understanding of existing neural approaches.

\subsubsection{Experimental Settings}
Our experimental environment is a desktop computer equipped with six NVIDIA GeForce GTX 1080Ti GPUs, Intel Xeon CPU, 24GB RAM, running on Ubuntu OS. For the neural approaches, we follow the implementation provided in their respective original papers~\cite{iyer2016summarizing,ahmad-etal-2020-transformer,wang-etal-2021-codet5}, and adopt the recommended hyperparameter settings~\cite{NeuralCodeSum,CodeT5}.

\begin{figure}[t]
  \centering
  \includegraphics[height=5cm,width=8.5cm]{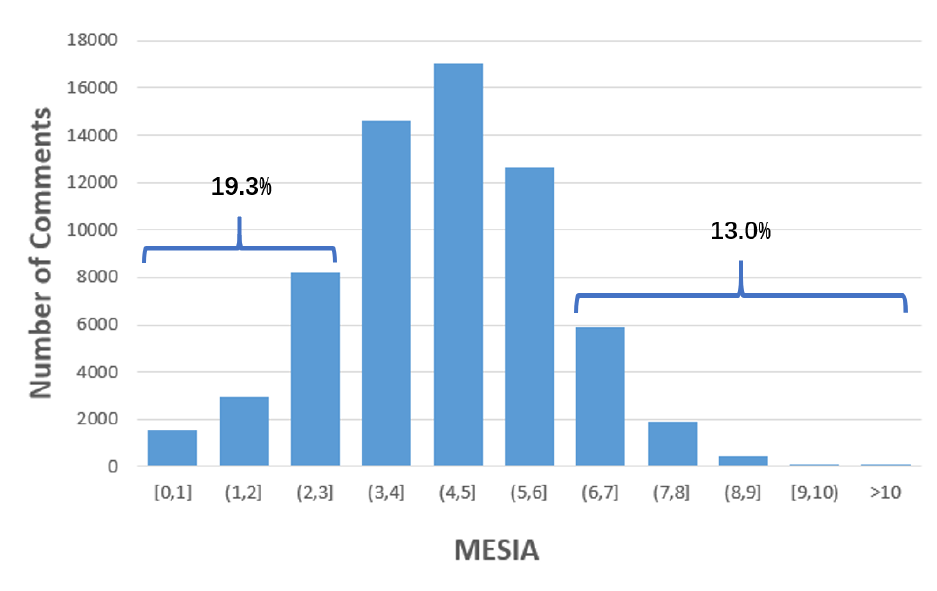}  
  \caption{Distribution of the MESIA value of the code comments in the TL-CodeSum dataset.}
  \label{fig:MESIA result}
  \Description{}
\end{figure}

\section{Study Result}
\label{sec:result}

In this section, we describe our study results and give some interesting findings.

\subsection{RQ1. How well does MESIA reflect the relative extent of supplementary information in method-level comments?}

\begin{figure*}[t]
  \centering
  \includegraphics[height=5.5cm,width=18cm]{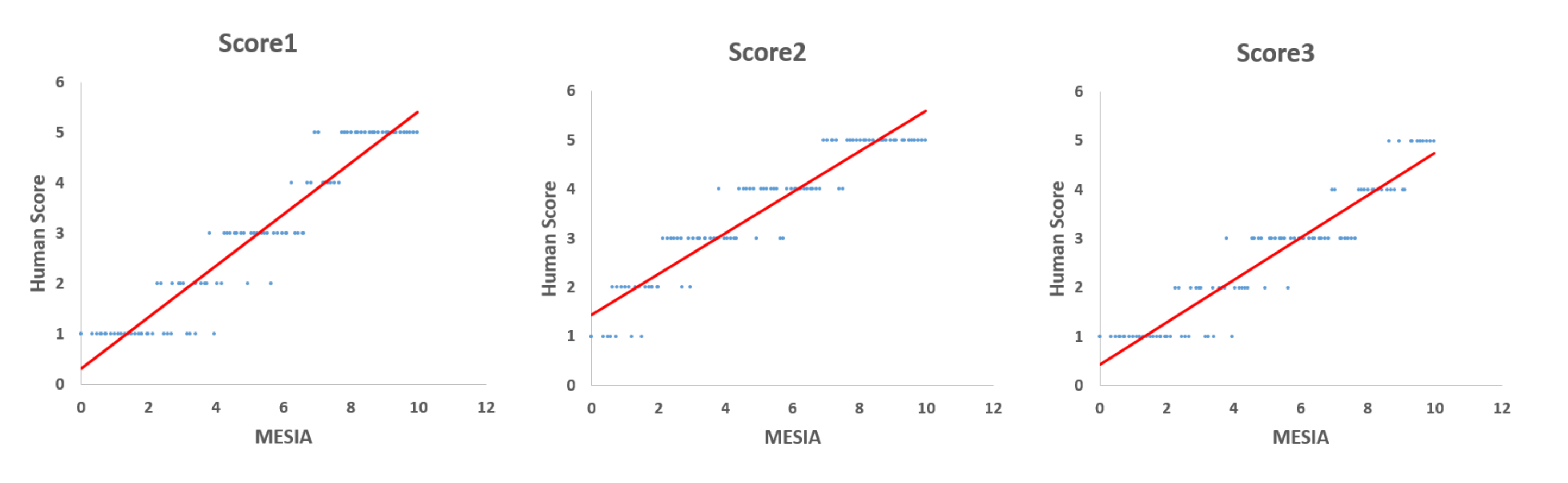} 
    \caption{Correlation between the MESIA values and the manual scores of each rater for the experimental code comments.
    MESIA and Score1 have a Spearman's $\rho$=0.9532 (p-value of 9.84e-53).
    MESIA and Score2 have a Spearman's $\rho$=0.9530 (p-value of 1.26e-42).
    MESIA and Score3 have a Spearman's $\rho$=0.9498 (p-value of 2.69e-51).
    We also calculate the Spearman's $\rho$ between the manual scores of different raters.
    Score1 and Score2 have a Spearman's $\rho$=0.9478 (p-value of 1.78e-50).
    Score1 and Score3 have a Spearman's $\rho$=0.9367 (p-value of 1.77e-46).
    Score2 and Score3 have a Spearman's $\rho$=0.9780 (p-value of 1.56e-68). All the results demonstrate a strong and
significant correlation.}
    \label{fig:humanStudy}
  \Description{}
\end{figure*}

To answer this research question, we first calculate the MESIA value for each code comment in the dataset. After calculation, we find that 99.8\% of the MESIA values are less than 10. Therefore, we divide the MESIA values into 11 intervals: [0,1],(1,2],...,(9,10], and (10,$+\infty$). Then, we analyze how many code comments there are in each interval of the MESIA value. Figure~\ref{fig:MESIA result} shows the results. We can see that most code comments get a MESIA value between 3 and 6. About 19.3\% of code comments get a small-MESIA value that is smaller than 3. About 13.0\% of code comments get a large-MESIA value that is larger than 6. 

\begin{figure}[t]
  \centering
  \includegraphics[height=4.5cm,width=8cm]{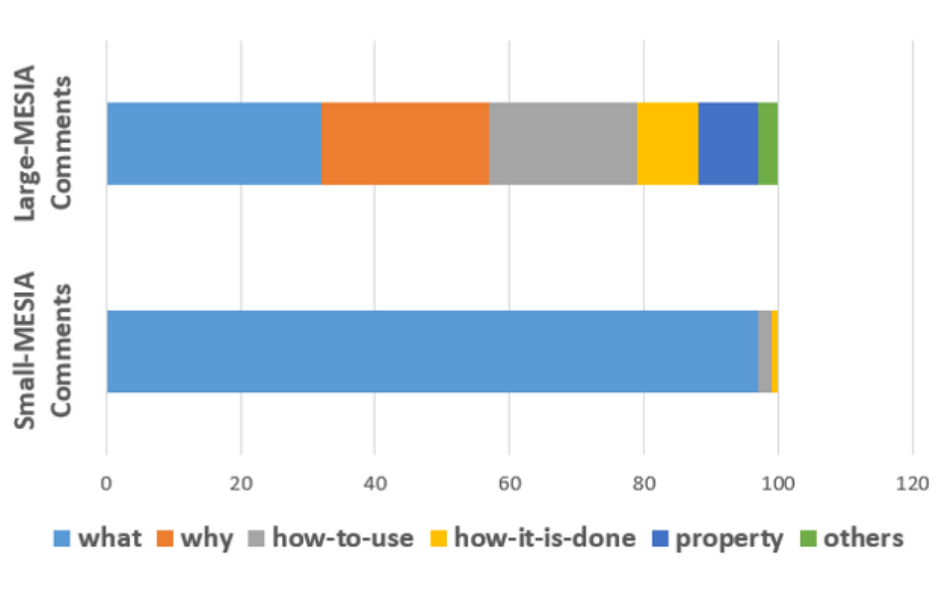}  
  \caption{Distribution of different kinds of code comments.}
  \label{fig:distribution}
  \Description{}
\end{figure}

To investigate whether MESIA is consistent with manual assessment of the relative supplementary extent of code comments, we conduct a manual experiment as follows. First, we construct 100 pairs of experimental data by randomly selecting 10 pairs of a method and its comment from each MESIA value interval\footnote{We merge code comments whose MESIA values are in (10,$+\infty$) with code comments whose MESIA values are in (9,10] because the amount of comments in these two intervals is too small.}. Second, we shuffle these 100 pairs of data and show them to three developers with more than five years of programming experience. We follow the hybrid mode of reading the method-level comment of a method and show only the method signatures and the comment to the developers. These three developers have never seen the data previously. Third, these three developers are asked to rate a score between 1 to 5 for each code comment independently according to their manual perception of the supplementary extent of the code comment. 1 represents that the comment brings little supplementary information and can even be left out. 2 represents that the comment brings a little supplementary information, which can still be derived by digesting the method signature. 3 represents that the comment brings substantive supplementary information to enhance the understanding of the method signature. 4-5 represents that the comment has brought supplementary information beyond the method signature. The higher score they rate, the larger extent of supplementary information they perceive the comment to have. Finally, we investigate the correlation between each rater's manual scores and the MESIA values for the experimental comments.

Figure~\ref{fig:humanStudy} presents the MESIA values and different raters' manual scores (described as Score1, Score2, and Score3, respectively) on the 100 pairs of experimental comments. We can see that although different raters may rate different scores for the same code comment (e.g., the second rater tends to give higher scores), the MESIA values of the comments are highly correlated with the manual scores given by each rater. We also calculate the Spearman's $\rho$~\cite{Spearman} of the MESIA values and the manual scores of each rater. The results indicate that MESIA has a strong and significant correlation with manual assessment of the relative supplementary extent of code comments.

\begin{figure}[t]
  \centering
  \includegraphics[height=3.5cm,width=\linewidth]{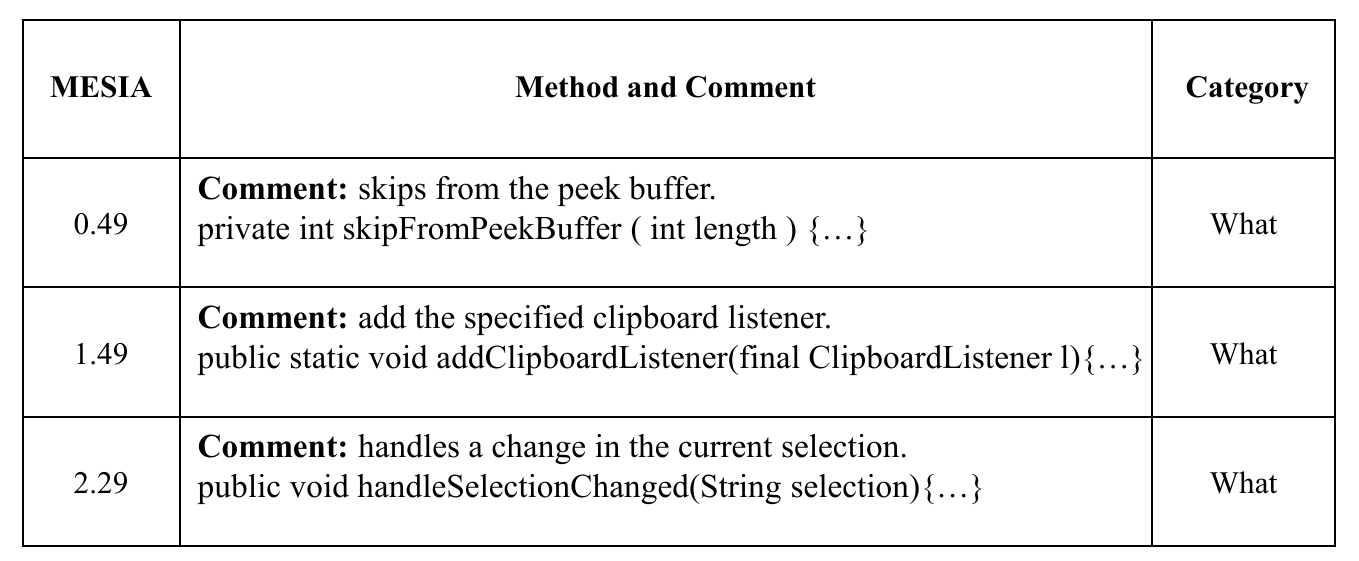} 
    \caption{Examples of small-MESIA comments.}
    \label{fig:lowercategory}
  \Description{}
\end{figure}

\begin{figure*}[t]
  \centering
  \includegraphics[height=8cm,width=\linewidth]{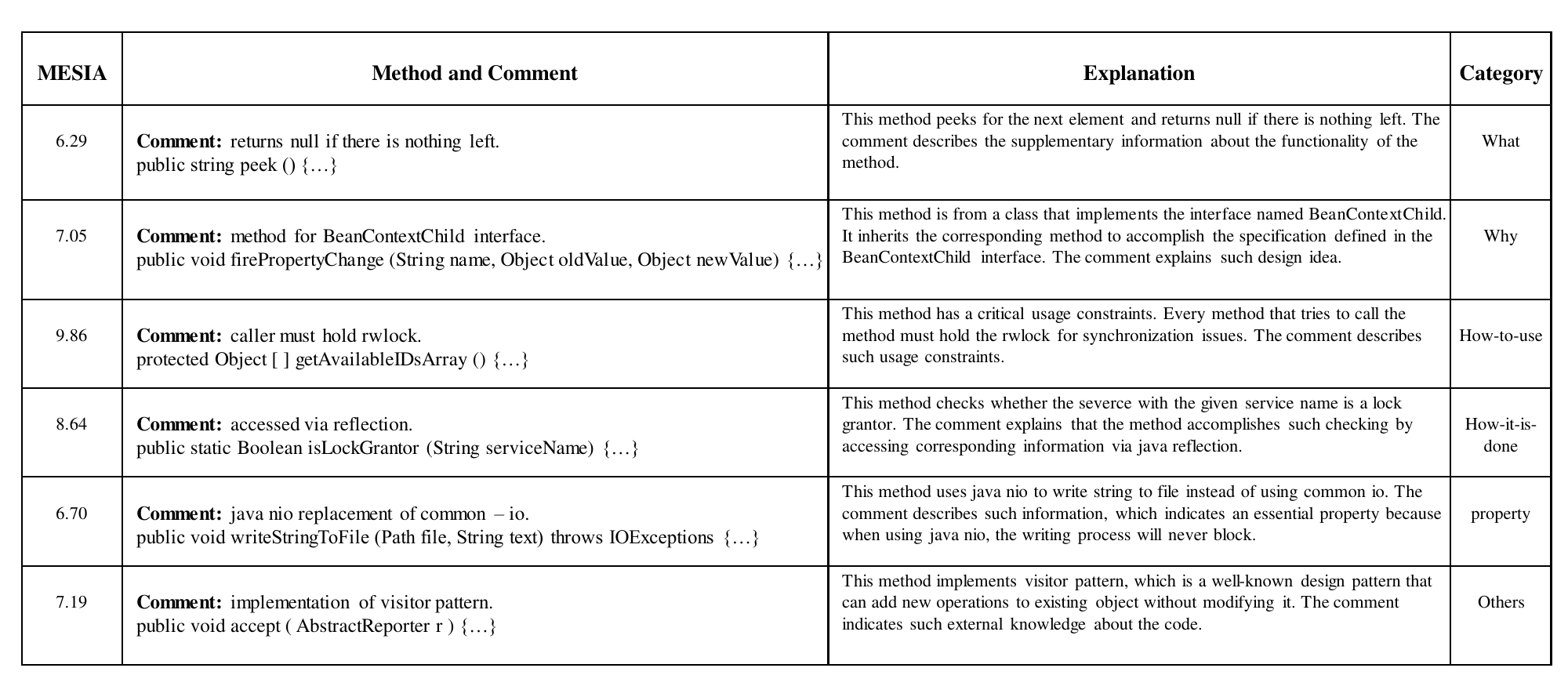} 
    \caption{Examples of large-MESIA comments.}
    \label{fig:highcategory}
  \Description{}
\end{figure*}

To investigate the difference between small-MESIA comments and large-MESIA comments, we further conduct a manual experiment. We randomly select 100 code comments with a small-MESIA value (\textless 3) and 100 code comments with a large-MESIA value (\textgreater 6), then ask the three developers to manually classify these code comments into different categories. We follow a clear taxonomy of code comments given by previous work~\cite{10.1145/3434280}. The taxonomy has six categories: (1) ``What" describes the functionality of the method. (2) ``Why" describes why the method exists. (3) ``How-to-use" describes how to use the method, including the method’s usage constraints. (4) ``How-it-is-done" describes the implementation details of the method. (5) ``Property" describes special properties (e.g., pre-conditions) of the method. (6) ``Others" describes other additional external knowledge about the method. Because each code comment in the dataset is only one sentence and tends to describe only a certain aspect of the method, we follow the previous work~\cite{10.1145/3434280} and ask the developers to classify these code comments into one of the six categories independently. If there is a disagreement between the classifying results, developers will discuss it and come to a consensus. Finally, for the small-MESIA comments, the Fleiss Kappa value~\cite{Fleiss1971MeasuringNS} of the classifying results is 0.88, which is almost a perfect agreement. For the large-MESIA comments, the Fleiss Kappa value of the classifying results is 0.76, which is also a substantial agreement.

Figure~\ref{fig:distribution} presents the classifying results of the code comments. For the small-MESIA comments, most comments fall into only the ``what" comment category (describing the functionality). Figure~\ref{fig:lowercategory} shows three small-MESIA comments. We can see that the content of these comments can be easily obtained from the method signature. In the hybrid mode of reading the method-level comment to comprehend a method,  the comment provides little supplementary information.
For the large-MESIA comments, they cover all the six comment categories. Figure~\ref{fig:highcategory} shows six large-MESIA comments in different categories. These comments are all real-world examples from the studied dataset, which has been denoised by previous work~\cite{10.1145/3540250.3549145}. After our manual checking, these comments are all relevant to their corresponding methods and accurate (see the explanation). We can see that the large-MESIA comments describe various kinds of essential information about methods. For example, the comment of the method \emph{``getAvailalbeIDsArray''} is \emph{``caller must hold rwlock''}, which is a vital constraint to use the method. The comment of the method \emph{``writeStringToFile''} is \emph{``java nio replacement of common-io''}, which indicates an important property about the method because when using java nio, the writing process will never block. With such code comments, developers can use the methods more safely and with more confidence.

\begin{result-rq}{Summary for RQ1}
MESIA is consistent with manual assessment of the relative supplementary extent of code comments. Small-MESIA comments occupy around 20\% of the dataset and mostly fall into only the WHAT comment category while large-MESIA comments can provide various kinds of essential information about methods.
\end{result-rq}

\subsection{RQ2. What is the capability of existing neural approaches to generate code comments with different MESIA values?}

As we find in RQ1, the existing dataset is a mixture of code comments with various extents of supplementary information. To answer this research question, we need to consider the MESIA values of the code comments in the test set.

We divide the test set into different test groups according to the MESIA values of the comments. The test set contains 5905 pairs of a method and its corresponding comment after denoising and deduplication. We rank these code comments according to their MESIA values and equally divide them into 10 test groups (G1 to G10). The average MESIA values of these 10 test groups are 1.51,2.70,3.25,3.68,4.07,4.45,4.85,5.27,5.83,7.03, respectively. 

\begin{figure}[t]
  \centering
  \includegraphics[height=6cm,width=7cm]{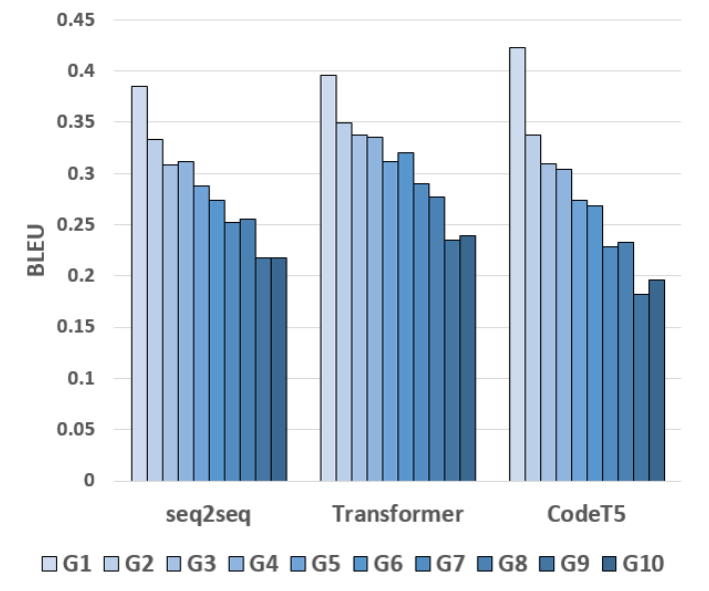} 
    \caption{Evaluation results of three models on different MESIA values of test groups.}
    \label{fig:RQ2}
  \Description{}
\end{figure}

\begin{figure*}[t]
  \centering
  \includegraphics[width =0.9\linewidth]{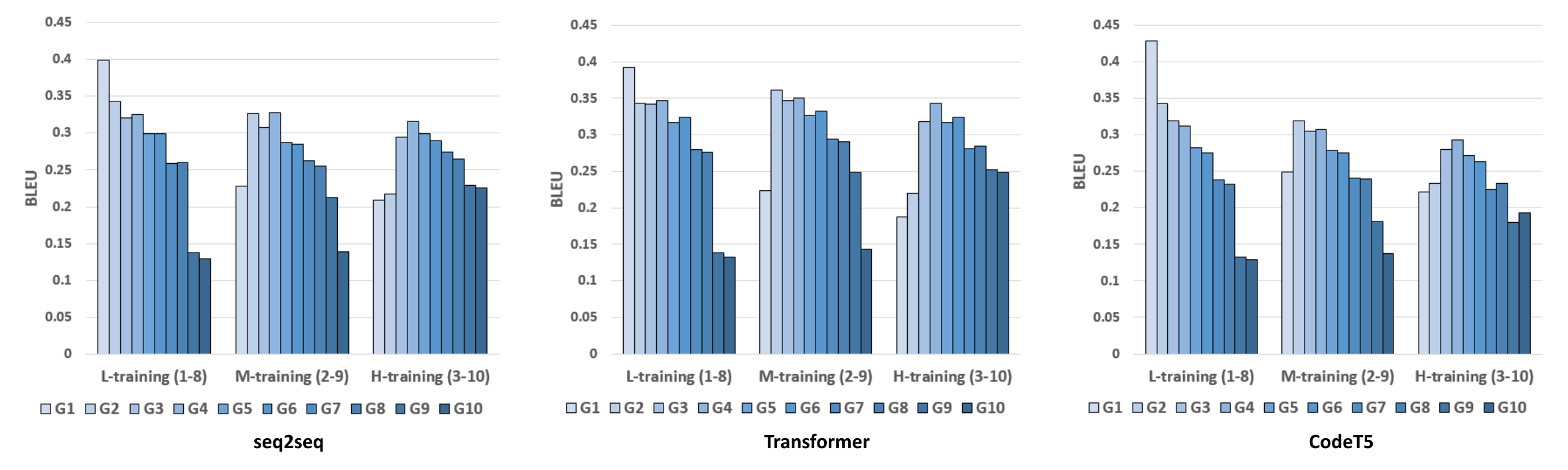}
  \caption{Evaluation results of three models on different MESIA values of test groups when using different training sets.}
  \label{fig:RQ3All}
  \Description{}
\end{figure*}

We train three neural models (seq2seq, Transformer, and CodeT5) on the training set and evaluate their effectiveness on each of the 10 test groups. Figure~\ref{fig:RQ2} shows the BLEU scores of the comments generated by the three models in the 10 test groups.
We can see that all models tend to achieve higher BLEU scores on test groups with smaller-MESIA comments. As the MESIA value of the reference comments in the test group increases, the effectiveness of all models gradually declines. Take the Transformer model as an example, in test group 1, the BLEU score can reach about 0.4; but in test group 10, the BLEU score is only 0.23. 

The results indicate that large-MESIA comments in the dataset are generally more difficult to generate than small-MESIA comments. We suspect that such difficulty may be caused by the following two factors. 1) Large-MESIA comments are more diverse as shown in RQ1. Therefore, the models are difficult to successfully generate the same category of code comments as the reference comments.
2) There are many small-MESIA comments in the training set, which influence the training of the models to pay more attention to the information in the method signature but overlook other essential information needed for generating the large-MESIA comments.

We suggest that the large-MESIA comments shall be given more attention in future research on code comment generation. If the comment-generation approaches fail to generate the small-MESIA comments, developers can still easily obtain the corresponding information from method signature in a hybrid mode of reading the method-level comment to comprehend a method; but if failing to generate the large-MESIA comments, developers may miss essential information needed to comprehend the method.

\begin{result-rq}{Summary for RQ2}
Based on the current dataset, existing neural approaches tend to get higher BLEU scores in generating the small-MESIA comments. The generation of large-MESIA comments has much room for improvement. 
\end{result-rq}

\subsection{RQ3. How well can MESIA be used to improve existing
neural approaches to generate large-MESIA comments?}

As shown in RQ2, the generation of large-MESIA comments has much room for improvement. In this research question, we want to investigate whether we can improve the existing neural approaches to generate large-MESIA comments by improving the training data with the MESIA metric. 

We construct three new training sets with the help of the MESIA metric. The training set originally contains 53506 code comments after denoising and deduplication. We rank these code comments according to their MESIA values and equally divide them into 10 groups. Every group contains 5350 pairs of a method and its comment (we omit the last six pairs for convenience). The average MESIA values from group 1 to group 10 are 1.56, 2.71, 3.28, 3.72, 4.12, 4.50, 4.89, 5.32, 5.87, and  7.07, respectively. The distribution is similar to the test groups. Then we construct three new training sets based on these 10 groups of data: the L-training set with groups 1-8 (average MESIA 3.76), the M-training set with groups 2-9 (average MESIA 4.30), and the H-training set with groups 3-10 (average MESIA 4.85). By using these training sets to train the neural models and evaluate the comment-generation result, we can investigate whether refining the training data with the MESIA metric can help existing approaches to generate more large-MESIA comments.

The construction of the three new training sets considers the following two factors. First, they shall contain the same size of sufficient training data. So we choose eight groups from the initial training data to construct the three new training sets and each new training set contains 42800 pairs of a method and its comment. Second, they shall contain a large amount of the same training data (groups 3-8). If the code comments in these training sets are all different, some other factors may also inevitably influence the effectiveness of the existing neural approaches. We need to minimize the influence of other factors.






We retrain the three neural models based on these three new training sets and evaluate their effectiveness again. The test set is also divided into 10 groups as RQ2. Figure~\ref{fig:RQ3All} presents the BLEU scores of the three models in the 10 test groups. We can see that based on the L-training set, all the approaches tend to achieve low BLEU scores on the large-MESIA test groups. As we gradually adjust the training set from L-training to H-training, the BLEU scores on the large-MESIA test groups gradually improve. The results indicate that different training data do affect the neural models' capability to generate large-MESIA comments. We can improve the generation of large-MESIA comments by reducing the proportion of small-MESIA comments in the training data.

\begin{figure}[t]
  \centering
  \includegraphics[height=6cm,width=8.5cm]{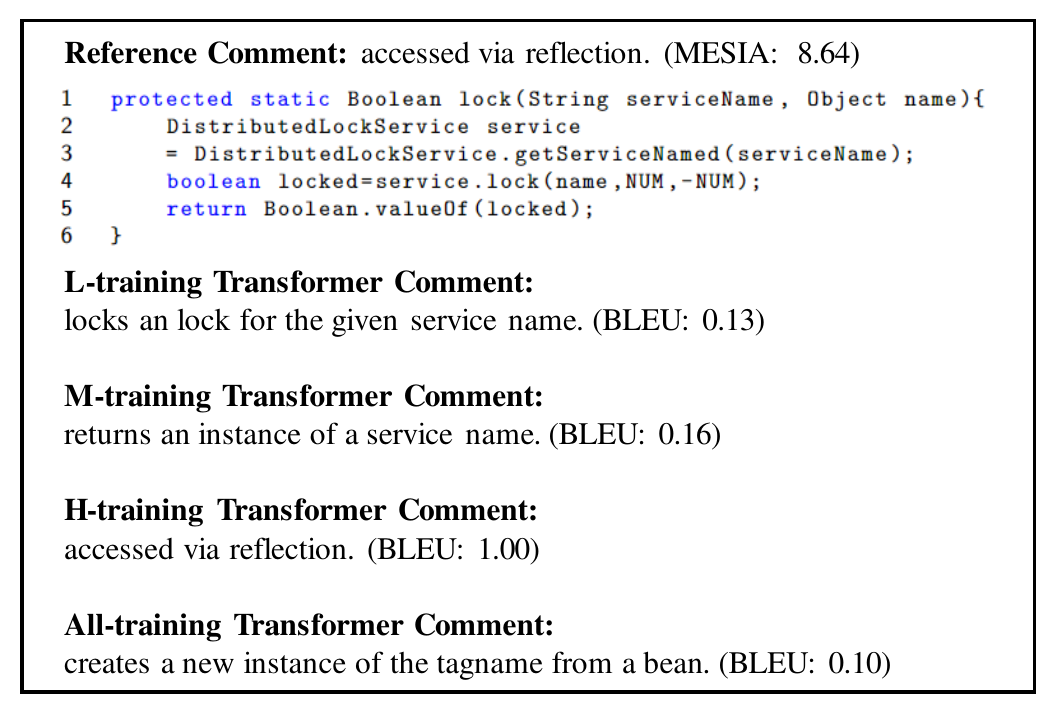}
  \caption{Comments generated for the method named ``lock'' under different MESIA values of training sets. }
  \label{fig:lock}
  \Description{}
\end{figure}

Figure~\ref{fig:lock} shows the comments generated for the method named ``lock''. In L-training, the generated comment is an intuitive summary that highly resembles the signature and has a small-MESIA value. In H-training, the generated comment is the same as the reference comment, which has a large-MESIA value and describes how the method accesses corresponding information to accomplish the locking (accessed via reflection). The comment generated under All-training (groups 1-10) is even wrong. The results indicate the usefulness and importance of the MESIA metrics, which can be leveraged to refine the training data to generate comments that bring more supplementary information for developers in a hybrid mode of reading the method-level comment to comprehend a method.

\begin{figure}[t]
  \centering
  \includegraphics[height=6cm,width=8.5cm]{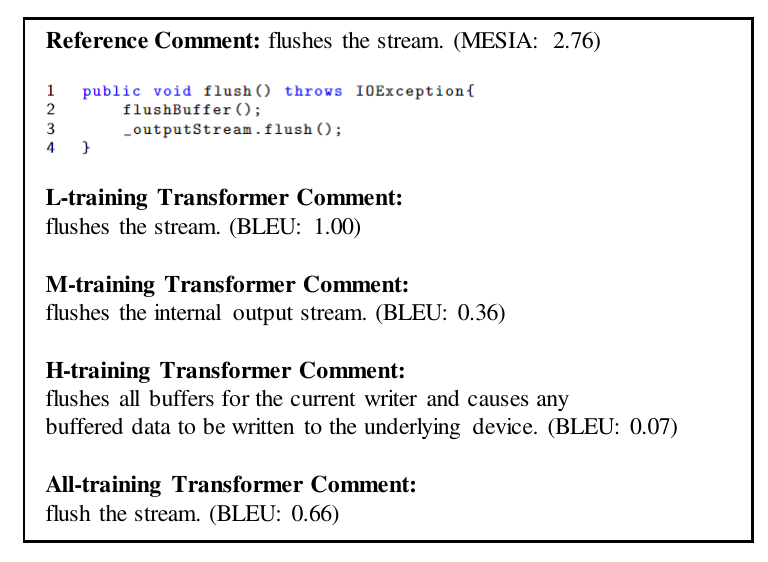}
  \caption{ Comments generated for the method named ``flush'' under different MESIA values of training sets. }
  \label{fig:flush}
  \Description{}
\end{figure}

For test sets with small-MESIA comments (e.g., G1 and G2), because they contain little supplementary information and H-training models tend to generate comments that bring more supplementary information not presented in the related reference comments, the BLEU score will naturally decline. For example, Figure~\ref{fig:flush} shows the comments generated for the method named \emph{``flush''}. In L-training, the generated comment is \emph{``flushes the stream''}, which is the same as the reference comment and has a perfect BLEU score but just brings a little supplementary information. In H-training, the generated comment is \emph{``flushes all buffers for the current writer and causes any buffered data to be written to the underlying device''}, which provides more supplementary information.  Such a comment has no appropriate ground truth for evaluation and gets a low BLEU score, but is actually pretty good. We search through the Internet and find that the method invokes two APIs in JDK (i.e., \emph{BufferedWriter.flushBuffer()} and \emph{OutputStream.flush()}). According to the JDK documentation~\cite{JDK}, the \emph{flushBuffer} API \emph{``flushes all buffers for the current writer''} and the \emph{flush} API \emph{``causes any buffered data to be written to the underlying device''}. The documentation confirms the accuracy of the generated comment. The model cannot generate such a comment even if it was trained in All-training, which contains all the data in original training sets.

The results reveal the influence of the supplementary nature of method-level comments on automatic comment generation. By reducing the proportion of small-MESIA comments in the training set, the retained model will be more effective in generating large-MESIA comments to better support developers' hybrid mode of reading the method-level comment to comprehend a method, but will get a lower BLEU score if evaluated in small-MESIA test sets. The results also indicate that with good training data, auto-generated comments can sometimes even surpass human-written reference comments. The latest study~\cite{2023arXiv230909558P} on artificial intelligence has also reported that summaries generated by existing popular large language models can surpass the reference summaries in the benchmark dataset. Having no appropriate ground truth for evaluation is an issue. Future work can explore human-like evaluation~\cite{2023arXiv230402554G} to better evaluate the automatic generation of code comments. 

\begin{result-rq}{Summary for RQ3}
We can improve the capability of existing neural approaches to generate large-MESIA comments by reducing the proportion of small-MESIA comments in the training set. The retrained model can generate meaningful large-MESIA comments that surpass the human-written reference comments for methods in the small-MESIA test set, but will get a lower BLEU score in evaluation because of having no appropriate ground truth. 
\end{result-rq}



\section{Discussion}
\label{sec:discussion}
In this section, we discuss our work's implications for future work and the threats to validity.

\subsection{Discussion of the MESIA Metric}

\subsubsection{\textbf{Significance of the MESIA metric.}} Several studies~\cite{9425926,10.1145/3510003.3510152} have observed developers' complaints about code comments' lack of supplementary information, and many developers advocate that "comments should not repeat the code"~\cite{stackOverflowCommentPractice} but "describe things that aren't obvious from the code"~\cite{PhilosophyOfSoftware}. Our proposed MESIA metric aims to evaluate the extent of supplementary information in code comments so as to improve the recent increasingly popular neural approaches for generating code comments. In practice, generic comments~\cite{10.1145/3510003.3510152} that do not provide much supplementary information are frequently encountered. Without the MESIA metric, we have no idea about the supplementary nature of code comments in the existing dataset for neural generation of code comments. Using the MESIA metric, we reveal the influence of the supplementary nature of method-level comments on automatic comment generation. We believe that our results can guide the usage of data in future work, and we encourage future work to apply our MESIA metric in more diverse datasets to validate the effectiveness in helping generate comments that better support developers' hybrid mode of reading the method-level comment to comprehend a method. 

\subsubsection{\textbf{Challenges of generating large-MESIA comments.}} There are still many challenges in generating large-MESIA comments. 
For example, the method named \emph{``writeStringToFile''} has a large-MESIA reference comment \emph{``java nio replacement of common - io''}, which contains external knowledge about the method. The Transformer model in H-training generates a comment  \emph{``writes a string to a file creating the file if it does not exist using the default encoding for the vm.''}, which is correct but still cannot provide the supplementary information in the reference comment. Based on our observation, it is common for the large-MESIA comment to contain supplementary information that cannot be directly derived from the associated method. Therefore, instead of just taking the given method as input, future work needs to consider more related information about the method, such as the file context or project context of the method, the documentation of the invoked APIs inside the method body, and other external programming knowledge about the method. In addition, because the large-MESIA comments can be diverse, it is a challenge to determine which kinds of supplementary information a given method may prefer or the developer may intend to write. Future work can explore approaches that can consider developers' intentions~\cite{10.1109/ICSE48619.2023.00073,2023arXiv230411384G} to generate comments that better satisfy developers' expectations. 

\subsubsection{\textbf{Improvements of the MESIA metrics}}
Future work can explore improving the MESIA metric in the following aspects. First, the calculation of the MESIA metric involves the probability of each word appearing in code comments. Currently, the probability is tied to the dataset. Therefore, the distribution of the words in the comment dataset may affect the MESIA metric. Future work can calculate the probability via a large scope to alleviate such an issue, or try some other ways of calculation such as the tf-idf. Second, to make the result of the metric comparable across different comment entries, we divide it by the comment length, which  may affect the MESIA metric. For some extremely short comments, although the quantity is small in the dataset, the MESIA value may be magnified. Future work can explore some smoothing approaches to alleviate such an issue. Third, although we have stemmed the words in method signature and words in code comments to a normalized form for comparison, future work may also need to further consider the abbreviations and synonyms to calculate the supplementary information amount better. Finally, the MESIA metric currently focuses on the scenario of comprehending and reusing methods in a hybrid mode. Therefore, we consider only the method signature in the metric. In other scenarios, future work needs to consider other additional aspects of the method, such as the invoked-methods' names inside the method body, to better assess the supplementary extent of code comments.


\subsection{Threats to Validity}
The threat to validity in this paper mainly includes the manual assessment process, the exploited dataset, and the exploited neural approaches for generating code comments.

The manual assessment process involves three developers and 300 pairs of a method and its comment. Although these three developers all have more than five years of programming experience, their perception of the supplementary extent of code comments may be subjective. Although these 300 pairs of a method and its comment are randomly chosen from different MESIA value intervals and are later shuffled, the amount of code comments  may still bring threats to validity. We plan to do the human study on more data with more experienced human participants to further study the MESIA metric.

The exploit dataset contains pairs of a method and its comment constructed by previous work~\cite{ijcai2018-314,10.1145/3540250.3549145}, and is further denoised and deduplicated. Although the dataset is widely used by later neural approaches for generating code comments, the results may not generalize to other datasets. We plan to conduct experiments on more datasets to further study the experimental results in the future.

The exploit neural approaches (seq2seq, Transformer, and CodeT5) are all from previous work~\cite{iyer2016summarizing,ahmad-etal-2020-transformer,wang-etal-2021-codet5}. Although these are three common types of approaches, more advanced approaches such as large language models remain to be studied. In this paper, we do not adopt the large language models such as ChatGPT~\cite{2023arXiv230512865S} because they have learned a large amount of data and the studied dataset (crawled from popular GitHub repositories long before November 2022) has probably been seen by these models and the evaluation will be affected. Nonetheless, our results have indicated that even for the traditional models, as long as trained with good training data, the auto-generated comments can sometimes surpass human-written reference comments. In the era of large language models, we believe that such a phenomenon will become more prevalent. We leave it as future work to comprehensively study the comment-generation results of large language models.

\section{Related Work} 
\label{sec:relatedwork}
Our work is mainly related to two lines of research: (1) Neural generation of code comments. (2) Code comment assessment. We discuss some of the related work in this section.
\subsection{Neural Generation of Code Comments }
In recent years, many neural approaches have been proposed for generating code comments~\cite{song2019survey,gros2020code}. The early work is mainly based on the seq2seq model. The representative work is codenn proposed by Iyer et al.~\cite{iyer2016summarizing}, which first introduce the seq2seq model from neural machine translation into code comment generation. Codenn treats the code snippet as a word sequence and translates it to a comment. Besides word sequence, some approaches also use the structural information of code, such as DeepCom proposed by Hu et al.~\cite{hu2018deep}, code2seq proposed by Alon et al.~\cite{alon2018code2seq}, ast-attendgru proposed by LeClair et al.~\cite{leclair2019neural}. Later, the Transformer model is applied in the comment generation task~\cite{9678882,ahmad-etal-2020-transformer,9401982}. For example, Ahmed et al.~\cite{ahmad-etal-2020-transformer} explore the Transformer model that uses a self-attention mechanism to effectively capture long-range dependencies when generating code comments. After that, some approaches~\cite{feng-etal-2020-codebert,guo2021graphcodebert,CodeGPT,wang-etal-2021-codet5} based on pre-training and fine-tuning are becoming popular. Recently, large language models~\cite{10.1145/3551349.3559548,10.1145/3551349.3559555,2023arXiv230512865S} demonstrate remarkable capability in code comment generation.

The preceding work has greatly promoted the automatic generation of code comments. Our work reveals the influence of the supplementary nature of method-level comments on automatic comment generation. By reducing the proportion of small-MESIA comments in the training data, we can generate comments that better support developers’ hybrid mode of reading the method-level comment to comprehend a method.


\subsection{Code Comment Assessment}
How to assess code comments has been discussed in several papers~\cite{10.5555/1894525.1894535,2021,10.1145/3510003.3510060,6613836,10.1016/j.jss.2022.111515}. The assessment approaches can be divided into two categories: manual assessment and automatic assessment.

For manual assessment, existing work mainly asks human participants to manually assess certain aspects of code comments, such as accuracy~\cite{sridhara2011generating,mcburney2014automatic}, adequacy~\cite{10.1109/ICSE48619.2023.00073}, and naturalness~\cite{iyer2016summarizing}. Manual assessment is important to ensure the quality of code comments but can be subjective and time-consuming.

For automatic assessment, there are three metrics widely used by existing neural approaches for evaluating the generated code comments:
BLEU~\cite{papineni2002bleu},
ROUGE\_L~\cite{lin2004rouge}, and
METEOR~\cite{banerjee2005meteor}.
Several studies~\cite{stapleton2020human,10.1145/3468264.3468588} have pointed out that these metrics cannot always reflect the quality of the generated code comments. For example, Stapleton et al.~\cite{stapleton2020human} do a human study and show that the BLEU and ROUGE\_L do not correlate with program comprehension. Roy et al.~\cite{10.1145/3468264.3468588} reassess these automatic assessment metrics and show that small changes in these metrics cannot guarantee human-perceivable improvement. We suspect that one important reason is that \textbf{these metrics rely on a reference comment to calculate a similarity for the generated comment in evaluation, but overlook the supplementary information of the reference comment}. For example, if the reference comment contains little supplementary information, even a generated comment that achieves the best evaluation result (e.g., BLEU = 1) brings little benefit, while a good supplementary comment will inevitably be undervalued. \textbf{Our proposed MESIA metric does not rely on a reference comment. Instead, MESIA aims to evaluate and guarantee the supplementary information for the reference comments. Therefore, MESIA can facilitate existing reference-based metrics for better evaluation.}

Some work~\cite{10.1007/s10664-014-9344-6,10.1145/3361242.3362699,1631117} has been proposed to understand and leverage the similarity between code and comments. However, the supplementary information of code comments is seldom studied. Within our knowledge, our proposed MESIA metric is the first attempt to assess the supplementary extent of code comments. Assessing such an aspect can better support developers’ hybrid mode of reading the method-level comment to comprehend a method.

\section{Conclusion}
\label{sec:conclusion}
In this paper, we have raised the awareness of developers' expectations for the supplementary information of method-level comments when reading the method-level comment for a method in a hybrid mode. We have proposed a new metric named MESIA to evaluate the extent of supplementary information that a code comment can provide beyond the method signature. With the MESIA metric, we have conducted experiments on a popular code-comment dataset and three common types of neural approaches to generate method-level comments. Our experimental results have revealed the influence of the supplementary nature of method-level comments on automatic comment generation. By reducing the proportion of small-MESIA comments in the training set, we can improve the capability of existing neural approaches to
generate large-MESIA comments. The generated comments can sometimes even surpass the human-written reference comments, and having no appropriate ground truth for evaluation is an issue that needs to be addressed by future work on automatic comment generation.

In future work, we plan to conduct experiments on more  datasets and more  neural approaches to further study the MESIA metric. We also plan to further improve neural approaches for generating large-MESIA comments. All of our data including the datasets, the source code, and the evaluation results are available at: \url{https://github.com/MESIA-CodeComment/MESIA}

\begin{acks}
This work is supported by National Key Research and Development
Program of China (Grant No. 2023YFB4503803), National Natural Science Foundation of China under Grant No. 62161146003, and the Tencent Foundation/XPLORER 
PRIZE.
\end{acks}

\balance
\bibliographystyle{ACM-Reference-Format}
\bibliography{reference.bib}
\end{document}